\documentstyle[amssymb,aps,multicol]{revtex}

\begin{document}
\preprint{}
\draft

\title{Control of open quantum system dynamics}

\author{Seth Lloyd${}^{\dagger}$ and Lorenza Viola${}^{}$}
\address{ d'Arbeloff Laboratory for Information Systems and
Technology, \\
Department of Mechanical Engineering,  %\\
Massachusetts Institute of
Technology, \\
Cambridge, Massachusetts 02139 }

\maketitle

\begin{abstract}
We investigate the control resources needed to effect arbitrary 
quantum dynamics. We show that the ability to perform measurements 
on a quantum system, combined with the ability to feed back the measurement 
results via coherent control, allows one to control the system to follow 
any desired open-system evolution. Such universal control can be achieved, 
in principle, through the repeated application of only two coherent
control operations and a simple ``Yes-No'' measurement.
\end{abstract}


\begin{multicols}{2}

Ever since the discovery of quantum mechanics more than a century
ago, the problem of controlling quantum systems has been an 
important experimental issue \cite{blaquiere:1987}.  
A variety of techniques are available for controlling quantum 
systems, and a detailed theory of quantum control has been developed
\cite{huang:1983,rabitz}.  Methods of geometric and coherent control are 
particularly powerful for controlling the coherent dynamics of quantum 
systems \cite{rabitz,brockett:1983}.  In coherent control, a series of
semiclassical potentials described by Hamiltonians $\{H_i\}$ are
applied to the system.  A basic result in coherent
control is that, for a finite dimensional quantum system, only a small 
set of Hamiltonians is required to enact any desired unitary 
transformation on the system \cite{huang:1983,rabitz}.  In general,
only two different Hamiltonians will suffice to achieve such universal 
control \cite{lloyd:1995}. Thus, arbitrarily complicated time evolutions 
can be built up out of simple building blocks. 

In this Letter we investigate the corresponding question for open-system,
incoherent dynamics.  Can an arbitrarily complicated open-system dynamics 
be built up by applying coherent control together with only a {\sl few} basic 
open-system operations?  Unlike the case of Hamiltonian dynamics, where complete
control can be attained purely in terms of open-loop manipulations as implied 
above, this turns out to be possible only if closed-loop control 
({\it i.e.}, feedback) is allowed.
We show that only one simple coupling to a measurement apparatus, 
combined with the ability to coherently manipulate the system and feed back 
the measurement results by the repeated application of a few basic control
operations, allows one to enact arbitrary open-system dynamics.

Let us first review the basic result of 
coherent control theory mentioned above. Consider an $n$-dimensional 
quantum system, $S$, 
%(all quantum systems with finite energy confined to a finite volume 
%are effectively finite dimensional). 
and suppose that one can apply potentials corresponding to a set 
of Hamiltonians $\{\pm H_i\}$, so that one can apply unitary operators 
of the form $e^{\pm i H_{i_n}t_n} \ldots e^{\pm H_{i_1}t_1}$.  Note that
$e^{iH_2\Delta t}e^{iH_1\Delta t}e^{-iH_2\Delta t}e^{-iH_1\Delta t}
= e^{[H_1,H_2]\Delta t^2} + O(\Delta t^3)$,
where $[H_1,H_2]$ is the commutator of $H_1$ and $H_2$.  
Repeating this procedure shows that applying Hamiltonians from the set allows 
one to enact any coherent evolution of the form $e^{-iAt}$, where $A$ belongs 
to the algebra generated from the set by commutation.  The time evolution is 
stroboscopic, matching up to the desired one at discrete intervals, 
and approximate.  But the error of approximation 
can be made as small as desired by making $\Delta t$ sufficiently small.  
Since typically only a small number of Hamiltonians (two, {\it e.g.}) are required 
to generate all possible Hamiltonians via commutation, arbitrary coherent
transformations can be built up from a small set of simple building blocks. 
This coherent control technique is open-loop because no information about the
state of the system is required to enforce the desired dynamics.

Now consider open-system dynamics.  
The simplest way to describe an open quantum system is to embed it in a 
closed quantum system.  Let the quantum system and environment be described 
by a joint density matrix $\rho_{SE}$. The time evolution of $\rho_{SE}$
is given by a unitary transformation $\rho_{SE} \mapsto U\rho_{SE} U^\dagger$.  
Before the transformation the system on its own is described by a reduced
density matrix $\rho_{S} = {\rm tr}_E \rho_{SE}$.  Afterward, its state is 
${\rm tr}_E  U\rho_{SE}U^\dagger$. 
%From the result of the previous paragraph, 
It is then immediately clear that if one has the ability to enact any 
coherent transformation on the system and environment taken together, then 
one can enact any desired open-system transformation of the system on its own
\cite{lloyd:1996}.

In general, however, one does not have control over the system's
environment, but only over the system itself.  If the system and the 
environment are initially uncorrelated, so that 
$\rho_{SE}(0)=\rho_S\otimes \rho_E$, the time 
evolution of $S$ can be described by a completely-positive, 
trace-preserving map $\rho_S \mapsto \sum_k A_k \rho_S A_k^\dagger$,
where $\sum_k A_k^\dagger A_k = \openone$ and the so-called Kraus 
operators $A_k$ can be derived from the embedding of the previous 
paragraph \cite{kraus:1983}.  
The most general infinitesimal limit of such a time evolution
is given by the Lindblad equation \cite{alicki:1987}:
\begin{eqnarray} 
{d \rho_S \over dt} & = & -i\, [H,\rho_S] \nonumber \\
& - & {1 \over 2}\sum_{i,j} a_{ij} 
\big( F_i^\dagger F_j \rho_S + \rho_S F_i^\dagger F_j -2 F_j 
\rho_S F^\dagger_i \big) \,, 
\label{lindblad1}
\end{eqnarray}
where $H=H^\dagger$ is the effective system Hamiltonian 
(the natural Hamiltonian possibly renormalized by a Lamb-shift term),
$\{ F_i \}$ is a basis for the space of bounded traceless 
operators on $S$, and ${\tt A}={\tt A}^\dagger=\{ a_{ij}\}$ is a positive 
semi-definite matrix.  After diagonalizing ${\tt A}$, Eq. (\ref{lindblad1}) 
can be cast in the equivalent canonical form 
\begin{eqnarray} 
{d\rho_S \over dt}& = & -i \, [H,\rho_S] \nonumber \\ 
& -& {1 \over 2} \sum_k \big( L_k^\dagger L_k \rho_S
+ \rho_S L_k^\dagger L_k -2 L_k \rho_S L^\dagger_k \big) \,,
\label{lindblad2}
\end{eqnarray}
for a set of bounded, traceless {\sl Lindblad operators} $\{L_k\}$ 
\cite{alicki:1987}. The Lindblad equation is a Markovian master equation that 
defines a quantum dynamical {\sl semigroup}: 
integrating the Lindblad equation over time $t$ defines an open-system 
transformation $\Lambda_t$ such that $\Lambda_t \Lambda_s =
\Lambda_{t+s}$ for $t,s\geq 0$.

Suppose that, in analogy with the infinitesimal open-loop prescription of 
the closed-system case, one attempts to build up arbitrary open-system
evolutions by applying coherent operations associated with Hamiltonians
$\{H_i\}$ together with some set of open-system infinitesimal operations
described by Lindblad operators $\{L_k\}$. Clearly, the implementation of the
coherent part of the dynamics, corresponding to $H$, poses no problem. 
However, one quickly arrives at an impasse while trying to enact the open-system
operation corresponding to an arbitrary ${\tt A}$, as the infinitesimal 
transformations $\Lambda_{\Delta t}$ are {\sl not} invertible in general.
%An arbitrary coherent (closed-system) time evolution can be built up from a few 
%basic building blocks by applying infinitesimal transformations in an open-loop 
%fashion as described above.  
%One might hope that one could build up arbitrary open-system
%time evolutions by applying coherent operations associated with Hamiltonians
%$\{H_i\}$ together with some set of open-system infinitesimal operations 
%described by Lindblad operators $\{L_k\}$.  Clearly, the coherent part of the 
%time evolution, corresponding to $H$, can be built up from coherent operations 
%as above. The difficulty lies in enacting the open-system operation
%corresponding to an arbitrary ${\tt A}$.  
%If one attempts to follow the infinitesimal prescription for building up 
%evolutions via commutation as in the closed-system 
%case, one quickly arrives at an impasse: the infinitesimal open-system 
%transformations $\Lambda_{\Delta t}$ are {\sl not} invertible. 
%Open-system operations of the form above comprise a semigroup, as opposite to the 
%group structure that underlies Hamiltonian dynamics. 
This has a drastic effect in determining the set of matrices ${\tt A}$ that 
can be reached by composition of infinitesimal open-loop transformations.
If we can apply infinitesimal operations corresponding to two different 
Lindblad equations specified by ${\tt A}$ and ${\tt A}'$, then we can generate
an infinitesimal operation corresponding to ${\tt A} + {\tt A}'$.    
But it is not in general possible to build up an arbitrary positive
semi-definite matrix by adding together a finite set of positive
semi-definite matrices.  
%The situation is analogous to that of density
%matrices, which are also positive semi-definite: adding together two
%density matrices creates a mixed density matrix.  And no matter how many
%density matrices one mixes together, one can never reach a pure state.   
Indeed, it has been shown recently that even in the case of a simple
two-state quantum system a {\sl continuous} set of distinct non-unitary 
open-loop transformations is required to generate an arbitrary Markovian 
dynamics \cite{kempe}.

%So in contrast to the coherent, closed-system case, in which only a few basic
%operations suffice to enact an arbitrary time evolution, no matter what the 
%dimension of the system's Hilbert space, an infinite repertoire of 
%independent open-loop transformations is required to build up an arbitrary 
%time evolution in the incoherent, open-system case. 

Turn now to closed-loop control.  In closed-loop control one
performs measurements on the system and feeds back the results of these
measurements by applying operations that are a function of the measurement 
results. Although both the system and the measurement apparatus are 
quantum-mechanical in principle, 
the feedback operations we consider here involve feeding back {\sl classical}, 
not quantum information. The case of feeding back quantum information 
via fully coherent quantum feedback was addressed in \cite{lloyd:1997}.  
Note that closed-loop control is intrinsically {\sl non}-Markovian in that the 
feedback loop retains memory of the system's state at previous times \cite{wiseman}.
As will now be shown, such non-Markovian feature is essential to allow for the 
generation of arbitrary open-system dynamics, including arbitrary Markovian 
Lindblad evolutions, based on a few simple operations.

To analyze closed-loop control, we need to describe quantum measurements.  
Quantum measurement is a special case of an open-system time evolution.  
%Not surprisingly, then, the most general picture of quantum measurement closely 
%resembles the general picture of open-system time evolution given above.
A generalized quantum measurement can always be described by letting the 
system to be measured interact with an auxiliary environment
(or ancilla) and then performing a conventional von Neumann measurement
on the latter \cite{peres}.  In terms of Kraus operators, a generalized 
measurement corresponds to a set of operators $A_{km}$ such that 
$\sum_{km} A^\dagger_{km} A_{km} =\openone$. Suppose that the state of $S$ 
is $\rho_S$ as above.  The measurement then gives the result $k$ with
probability $p_k = {\rm tr}\sum_m A_{km} \rho_S A^\dagger_{km}$,
leaving $S$ in the state $\rho_k=\sum_m A_{km} \rho_S A^\dagger_{km}/p_k$.  
The information that can be gathered about a quantum system is maximized for 
so-called ``pure'' measurements \cite{peres,barnum}, in which case a single value 
value of $m$ occurs. 
%In this case the measurement gives the result $k$ with 
%probability $p_k = {\rm tr} A^\dagger_k A_k \rho_S $, leaving $S$ in the state 
%$A_k \rho_S A^\dagger_k/p_k$. 
We will restrict to pure measurements henceforth.

A particularly interesting class of pure measurements occurs when the 
Kraus operators are positive (hence Hermitian). Such measurements, which 
can be realized by coupling the system to a suitable ``pointer variable'' 
of the measuring apparatus, are the ``least disturbing'' measurements that 
can be effected on $S$ \cite{barnum}. As such, they supply a natural analogue 
of von Neumann's model of projective measurements. 
%Suppose that one can realize such a generalized measurement corresponding 
%to an arbitrary set of positive measurement operators, and feed back
%the results of this measurement by applying coherent control.  
Suppose that one can perform such a generalized measurement corresponding to a 
set of positive operators $B_k$, and feed back the results by applying a
coherent transformation $U_k$ when the $k$-th outcome is found.  The net effect 
on the system is to apply the transformation
\begin{equation}
\rho_S \mapsto \sum_k U_k B_k \rho_S B^\dagger_k U_k^\dagger\,.
\label{polar} 
\end{equation}
That is, feedback gives rise to an open-system operation
corresponding to Kraus operators $A_k= U_k B_k$.
But, from the polar decomposition theorem, any operator can be written 
$A_k = U_k |A_k|$, where $|A_k| = \sqrt{A^\dagger_k A_k}$ is positive  
and $U_k$ is defined to be $\openone$ on the kernel ${\cal K}$ of $A_k$ 
%(the subspace of states $|\chi\rangle$ such that $A_k |\chi\rangle =0$) 
and $A_k |A_k|^{-1}$ on the orthogonal subspace ${\cal K}^\perp$ 
\cite{handbook}.  
As a consequence, the ability to make a generalized measurement corresponding 
to an arbitrary positive operator ($|A_k|$) together with the ability to 
apply an arbitrary coherent unitary transformation conditioned on the 
result of that measurement ($U_k$), allows one to enact an arbitrary 
open-system operation ($A_k = U_k |A_k|$). 

The polar decomposition provides a natural separation of a general quantum 
operation in terms of a measurement step, followed by a feedback step.  
This decomposition, which has been implicitly used throughout the history of 
quantum measurement \cite{blaquiere:1987,kraus:1983,barnum}, 
has recently proven useful to investigate the trade-off between information and 
disturbance in quantum feedback control \cite{doherty}.  In our context, 
since we have assumed the ability to build up arbitrary coherent transformations, 
we need only to find a way to perform measurements corresponding to arbitrary 
positive operators in order to enact an arbitrary open-system time evolution.  

Consider the simplest possible example of a quantum measurement, a so-called
``Yes-No'' measurement \cite{kraus:1983} in which the measurement apparatus $M$ 
consists of a single two-level quantum system with states 
$|0\rangle$, $|1\rangle$.
Assume that  $\rho_M = |0\rangle\langle 0|$ initially, and couple the system 
to the apparatus via an interaction Hamiltonian 
$H=\gamma X\otimes Y$, where $\gamma > 0$ is a coupling constant, and $X$, $Y$ 
are Hermitian operators acting on $S$, $M$  respectively.
We choose $X$ to be a positive operator projecting onto a pure state of $S$, 
and let $Y=|0\rangle \langle 1| + |1\rangle\langle 0|=\sigma_x$.
The state of the system and apparatus after they interact for a time $t$ is
\begin{eqnarray} 
%e^{-i\gamma tX\sigma_x} \rho_S\otimes |0\rangle\langle 0|
%e^{+i\gamma tX\sigma_x} \nonumber \\ 
\rho_{SM}(t) & = &  \big(\cos(\gamma t X) -i\sin(\gamma t X)\otimes \sigma_x \big)
\rho_S \otimes \rho_M \nonumber \\  
& \mbox{} & \big(\cos(\gamma t X) + i\sin(\gamma t X)\otimes \sigma_x \big)\,.
\label{finite}
\end{eqnarray}
Now make a von Neumann measurement of $0,1$ on the ancilla. 
The result $0$ occurs with probability 
$p_0={\rm tr} \cos^2(\gamma t X)\rho_S$, in which case $S$ is in the state 
$\rho_0= \cos(\gamma t X)\rho_S \cos(\gamma t X)/p_0$. The result $1$ occurs 
with probability $p_1 = {\rm tr} \sin^2(\gamma t X)\rho_S$, leaving $S$ in the
state $\rho_1 = \sin(\gamma t X)\rho_S\sin(\gamma t X)/p_1$.  
In other words, this Yes-No measurement corresponds to Hermitian Kraus
operators $\cos(\gamma t X), \sin(\gamma t X)$.

The ability to perform arbitrary coherent transformations on the system
can now be used to transform this simple Yes-No measurement 
into an arbitrary Yes-No measurement. This can be accomplished by applying an 
average Hamiltonian technique \cite{viola2} {\sl during} the system's coupling 
to the auxiliary quantum system in the course of the measurement.
%While the system is coupling to the auxiliary system, and 
Before the von Neumann measurement is made on the ancilla, perform the following 
sequence of rapid coherent transformations on the system:
Apply a unitary transformation $V_1$, wait for a time $\Delta_1$, then perform 
the transformation $V_1^\dagger$; now apply a second transformation $V_2$, wait 
for a time $\Delta_2$, and perform the transformation $V_2^\dagger$; {\it etc.}, 
for $N$ periods.  Assume that the $V_i$'s are effected on a time scale short 
compared with the $\Delta_i$, and let $\sum_i \Delta_i = \Delta t$.  
Repeat $N=t/\Delta t$ times.  If $N$ is large, then to lowest order in 
$\Delta t$ the net effect of these repeated transformations is to replace $X$ in 
Eq. (\ref{finite}) by $\overline{X} = \sum_i (\Delta_i/\Delta t) V^\dagger_i X V_i$. 

Starting from a given projector $X$ onto a pure state, 
%so that the eigenvalues of $X$ are $1,0,\ldots,0$. 
any positive operator with unit trace can be written in the form 
$\overline{X}$ for some $V_i$, $\Delta_i \geq 0$. But with an appropriate 
choice of the operator $\overline{X}$, and the parameters 
$\gamma$, $t$, any pair of positive Kraus operators can be represented in the 
form $B_0=\cos(\gamma t\overline{X})$, $B_1=\sin(\gamma t\overline{X})$.
So our simple measurement procedure together with the ability to perform 
coherent control translates into the ability to make arbitrary two-outcome 
minimally disturbing measurements.

Now add feedback.
Performing the von Neumann measurement on the ancilla and feeding back 
the result of of the measurement by applying $U_0$ if the result is $0$ and 
$U_1$ if the result is $1$ gives an open-system time evolution
$\rho_S \mapsto A_0 \rho_S A^\dagger_0 + A_1 \rho_S A_1^\dagger$, 
where $A_0 = U_0 \cos(\gamma t \overline{X})$  and 
$A_1 = U_1 \sin(\gamma t \overline{X})$. But, by polar decomposition, any set 
of two Kraus operators can be written in this form.  
As a result, the ability to perform coherent control combined with the 
ability to perform a single, simple measurement on the system translates 
into the ability to enact an arbitrary two-operator open-system time evolution.  

Generalized measurements with more than two outcomes can be constructed in 
a straightforward way \cite{kraus:1983}.  For example, if one can make 
measurements corresponding to arbitrary $A_0,A_1$, a measurement corresponding 
to arbitrary Hermitian $B_0, B_1, B_2$, with $B_0^2+B_1^2+B_2^2=\openone$, can 
be enacted as follows.  
Perform a measurement corresponding to $B_0, B'_1 = \sqrt{B_1^2 + B_2^2}$.  
If the result of this measurement is $0$, do nothing.  If the result of
this measurement is $1$, perform a second measurement
corresponding to $A_0 = B_1 {B'_1}^{-1}$, $A_1 = B_2 {B'_1}^{-1}$ 
(note that ${B'_1}^{-1}$ is well-defined on 
the set of states that result when the first measurement has outcome $1$). 
This measurement is generally not Hermitian, but the previous paragraph shows 
that we can do arbitrary generalized measurements with two outcomes.
The outcome $0$ then corresponds to $B_0$, the outcome
$10$ corresponds to $B_1$, and the outcome $11$ corresponds
to $B_2$.  Arbitrary measurements with multiple (possibly non-Hermitian) outcomes
can be built up by allowing feedback and by following an analogous procedure.

To summarize, the ability to perform a {\sl single} simple measurement on 
the system, together with the ability to apply coherent control to feed back the 
measurement results, allows one to enact an arbitrary finite-time open-system
evolution $\rho_S \mapsto \sum_k A_k \rho_S A_k^\dagger$. 

Now turn to the case of infinitesimal open-system time evolutions
governed by the Lindblad equation.
To address infinitesimal time evolutions, we can imagine that the 
interaction between the system and the ancilla representing the 
measurement apparatus only takes place for a short period of time. 
Accordingly, the unitary propagator $\exp(-iHt)$ evolving the system and 
apparatus together can be expanded to second order in time.
%In the presence of a generic infinitesimal interaction $H$,
%the initial state of the system and apparatus, 
%$\rho_{SM}(0)=\rho_S\otimes \rho_M$, evolves to
%$e^{-iHt} \rho_S\otimes \rho_M e^{+iHt}$.  
%Expanding this equation to second order in $t$ gives
%\begin{equation}
%\rho_{SM}(t) = \rho_{SM} -i t [H, \rho_{SM}]
%-(t^2/2)(H^2 \rho_{SM} - 2 H\rho_{SM}H + \rho_{SM} H^2)\,.
%\label{infinitesimal}
%\end{equation}
By writing $H=\gamma \, X\otimes Y$ as above, one gets 
\begin{eqnarray}
\rho_{SM}(t) & = & \rho_S\otimes\rho_M
-i\gamma t (X \rho_S\otimes Y\rho_M - \rho_S X \otimes \rho_M Y)
\nonumber \\
& - & {\gamma^2 t^2 \over 2}\, \big(X^2 \rho_{S}\otimes Y^2\rho_M
- 2 X\rho_{S}X \otimes Y\rho_M Y \nonumber \\
& + & \rho_{S} X^2 \otimes \rho_M Y^2 \big)\,.
\label{pointer}
\end{eqnarray}
The time evolution for the system on its own is obtained by tracing over 
the measurement apparatus and taking the small-time limit ({\it i.e.}, $t$
small but finite \cite{note}). This results in the Lindblad equation for 
the system,
\begin{equation}
{ d\rho_S \over dt} = -ia\, [X,\rho_S] 
- b\, (X^2\rho_S - 2 X \rho_S X + \rho_S X^2),
\label{measurement}
\end{equation}
where $a=\gamma \,{\rm tr}_M Y \rho_M=0$ and 
$b=(\gamma^2 t/2)\,{\rm tr}_M Y^2 \rho_M$ = $\gamma^2 t/2$ for 
$\rho_M=|0\rangle \langle 0|$, $Y=\sigma_x$. That is, comparing with 
(\ref{lindblad2}), the simple coupling above results in a single Lindblad 
operator $L=\sqrt{2b}X$.

To complete the infinitesimal measurement and feed back the classical
information, continue just as in the non-infinitesimal case, by making a 
von Neumann measurement on the auxiliary system. 
If the result is $0$, do nothing; if the result is $1$, apply the coherent 
transformation $U$ to $S$. By using Eq. (\ref{pointer}) and tracing over $M$, 
%\begin{eqnarray}
%\rho_{SM}(t) = &\rho_S\otimes |0\rangle\langle 0|
%-i\gamma t (X \rho_S\otimes |1\rangle\langle0| 
%- \rho_S X \otimes |0\rangle\langle 1| ) \nonumber \\
%& -(\gamma^2t^2/2)(X^2 \rho_{S}\otimes |0\rangle\langle 0| 
%- 2 X\rho_{S}X \otimes |1\rangle \langle 1|
%+ \rho_{S} X^2 \otimes |0\rangle\langle 0|)\,.
%\label{qubit}
%\end{eqnarray}
the resulting state of the system is
\begin{equation}
\rho_{S}(t) = \rho_S - {\gamma^2 t^2 \over 2}\, 
\big( X^2 \rho_{S} - 2 UX\rho_{S}XU^\dagger 
+ \rho_{S} X^2 \big)\,.
\label{qubit2}
\end{equation}
In the small-time limit, this gives rise to a Lindblad equation 
(\ref{lindblad2}) with $L=\sqrt{2b}\,UX$.
%\begin{equation}
%{d\rho_S\over dt} =  
%- b (L^\dagger L\rho_S - 2 L \rho_S L^\dagger + \rho_S L^\dagger L),
%\label{measurement3}
%\end{equation}
That is, feeding back the result of the infinitesimal measurement 
yields a Lindblad equation 
%with no Hamiltonian term (the measurement has eliminated this term 
%via the quantum Zeno effect) 
with a single Lindblad operator proportional to $UX$, where $X$ is a 
unit-trace positive operator and $U$ is of our choosing.

The average Hamiltonian techniques described above can now be used
to construct an infinitesimal measurement corresponding to an arbitrary positive
operator $\overline{X}$.  Here, care must be taken to insure that the sequence 
of average Hamiltonian operations $V_i$ can be performed within the small 
``coarse-graining'' time $t$ above.  
Note that what is important in the derivation of the Lindblad equation 
(\ref{measurement}) is not that $t$ be small but rather that $\gamma t$ be small: 
accordingly, even though the relevant control time scale $\Delta t$ is not 
vanishingly small, $\gamma$ can always be made sufficiently small that the 
averaging operations can be fit within a time $t$ for which the derivation holds. 
Feeding back the results of the measurement allows one to enact a Lindblad 
equation governed by an arbitrary single Lindblad operator 
$L = \sqrt{2b}\, U \overline{X}$. 

%By performing average Hamiltonian techniques during the measurement process, 
%then feeding back the results of the measurement, one can enact any desired 
%Lindblad equation with a single Lindblad operator $L$. 
The same conclusion could have been also established directly from the ability 
to implement arbitrary two-operator open-system evolutions. 
In fact, in the small-time limit discussed above, the quantum operation 
specified by 
$A_0=U_0 \cos(\gamma t \overline{X})$ $\simeq \openone - 
(\gamma^2 t^2/2) \, \overline{X}^2$, 
$A_1=U_1 \sin(\gamma t \overline{X})$ $\simeq \gamma t \, U_1 \overline{X}$, 
with $U_0=\openone, U_1=U$, is formally equivalent to the Lindblad equation 
(\ref{lindblad2}) with $L=\sqrt{2b}\, U\overline{X}$. 
In terms of the matrix {\tt A} appearing in (\ref{lindblad1}), the ability to 
generate an arbitrary Lindblad operator $L=\sum_i c_i F_i$, $c_i \in {\bf C}$, 
translates into the ability to generate any rank-one matrix 
${\tt A}=\{ c_i c_j^\ast \}$ \cite{kempe}.  
As noted above, Lindblad equations with multiple $L_k$ can be enacted by 
performing the operations corresponding to each of the $L_k$ in succession.  
So by making a simple infinitesimal measurement, manipulating it by average 
Hamiltonian techniques, and coherently feeding back the results of the 
measurement, one can enact in principle an arbitrary Lindblad equation.  
The enactment is scaled in time, stroboscopic, and approximate, but the scaled 
time interval of the stroboscopic evolution and the error in the approximation 
can be made as small as possible by increasing the number of 
control operations per step.  

In conclusion, coherent control combined with feedback allows one to enact 
an arbitrary open quantum system dynamics.  In contrast to the case
of closed quantum systems, in which an arbitrary Hamiltonian time evolution 
can be enacted using only a few open-loop operations, the open-loop control 
of open-system dynamics requires in general an infinite repertoire of 
independent operations.  But closing the loop and feeding back the results of 
the measurements allows any desired open-system transformation to be enacted 
using only the coherent tools required for Hamiltonian control 
together with a single, simple quantum measurement. 
The ability to implement a similar control strategy in the appropriate 
small-time limit allows the construction of any desired continuous-time 
Markovian evolution described by a Lindblad master equation as well.

\vspace*{1mm}

This work was supported by DARPA/ARO under the QUIC initiative. \\
${}^\dagger$ Corresponding author: {\tt slloyd@mit.edu}

%\vspace*{-5mm}

\end{multicols}

\begin{references}

%\vspace*{-5mm}

%\bibitem{butkovskiy} A. Butkovskiy and Y. Samoilenko
%{\em Control of Quantum-Mechanical Processes and Systems} 
%(Kluwer Academic, Dordrecht, 1990).

\bibitem{blaquiere:1987} {\em Information Complexity and Control in 
Quantum Physics}, edited by A. Blaquiere, S. Diner, G. Lochak 
(Springer, New York, 1987). 

\bibitem{huang:1983} G. Huang, T. Tarn, and J. Clark, J. Math. Phys. 
{\bf 24}, 2608 (1983).

\bibitem{rabitz} A. Peirce, M. Dahleh, and H. Rabitz, Phys. Rev. A 
{\bf 37}, 4950 (1988); M. Dahleh,  A. Peirce, and H. Rabitz, {\em ibid.}
{\bf 42}, 1065 (1990); W. Warren, H. Rabitz, and M. Dahleh, Science 
{\bf 259}, 1581 (1993); V. Ramakrishna {\em et al.}, {\em ibid.} {\bf 51},
960 (1995).

\bibitem{brockett:1983} R. W. Brockett, SIAM J. Control {\bf 10}, 265 (1972); 
{\em Differential Geometric Control Theory}, edited by Brockett, R. Millman, 
and H. Sussman (Birkhauser, Boston, 1983).

\bibitem{lloyd:1995} S. Lloyd,  Phys. Rev. Lett. {\bf 75}, 346 (1995).

\bibitem{lloyd:1996} S. Lloyd, Science {\bf 273}, 1073 (1996).

\bibitem{kraus:1983} K. Kraus, {\em States, Effects, and Operations} 
%Fundamental Notions of Quantum Theory} 
(Springer, Berlin, 1983). 

\bibitem{alicki:1987} 
R. Alicki and K. Lendi, {\em Quantum Dynamical Semigroups and Applications}
(Springer-Verlag, Berlin, 1987).

\bibitem{kempe} D. Bacon {\it et al.}, {\sc lanl} e-print 
{\tt quant-ph/0008070}.

%\bibitem{viola} L. Viola and S. Lloyd, Phys. Rev. A {\bf 58}, 2733 (1998).

\bibitem{lloyd:1997} S. Lloyd, Phys. Rev. A {\bf 62}, 022108 (2000).

\bibitem{wiseman} However, closed-loop control can give rise to an 
effectively Markovian dynamics for the system considered on its own in the 
limit where the controller only retains memory of a single time step. See 
H. Wiseman and G. J. Milburn, Phys. Rev. Lett. 
{\bf 70}, 548 (1993); H. Wiseman, Phys. Rev. A {\bf 49}, 2133 (1994).

\bibitem{peres} A. Peres, {\em Quantum Theory: Concepts and Methods} 
(Kluwer Academic, Dordrecht, 1993).

%\bibitem{nielsen:1997} M. A. Nielsen {\it et al.}, {\sc lanl} e-print 
%{\tt quant-ph/9706064}.

\bibitem{barnum} H. N. Barnum, Ph.D. Thesis, University of New Mexico (1998),
available at {\tt http://wwwcas.phys.unm.edu/} \~{\tt hbarnum/homepage.html}.

\bibitem{handbook} H. L\"{u}tkenpohl, {\em Handbook of Matrices}
(John Wiley \& Sons, New York, 1996).

%\bibitem{braginsky} V. B. Braginsky and F. Y. Khalili, 
%{\em Quantum Measurement}
%(Cambridge University Press, Cambride, 1992). 

\bibitem{doherty} A. C. Doherty, K. Jacobs, and G. Jungman, 
{\sc lanl} e-print {\tt quant-ph/0006013}.

\bibitem{viola2}
L. Viola, E. Knill, and S. Lloyd, Phys. Rev. Lett. {\bf 82}, 2417 (1999); 
L. Viola, S. Lloyd, and E. Knill, Phys. Rev. Lett. {\bf 83}, 4888 (1999). 

\bibitem{note} Physically, the small time $t$ can be regarded as a coarse-graining
time resulting from the memory time scale over which the auxiliary system resets
its state.

\end{references}
\end{document}